\documentclass[showpacs,floatfix,twocolumn]{revtex4}

\usepackage{graphicx}

\def\gfo{$\gamma$-Fe$_2$O$_3$}
\def\mfo{MnFe$_2$O$_4$}

\begin{document}

\title{Positive and negative magnetocapacitance in magnetic nanoparticle
systems} 

\author{G. Lawes$^{1,2}$, R. Tackett$^{1}$,O. Masala$^{3}$, B.
Adhikary$^{1}$, R. Naik$^{1}$, and  R. Seshadri$^{3}$} 

\affiliation{$^1$
Department of Physics and Astronomy, Wayne State University, Detroit, MI 48201}
\affiliation{$^2$ Los Alamos National Laboratory, Los Alamos, New Mexico 87545}
\affiliation{$^3$ Materials Department and Materials Research Laboratory, 
University of California, Santa Barbara, CA 93106}

\date{\today}

\begin{abstract} The dielectric properties of MnFe$_2$O$_4$ and
$\gamma$-Fe$_2$O$_3$ magnetic nanoparticles embedded in insulating matrices
were investigated. The samples showed frequency dependent dielectric anomalies
coincident with the magnetic blocking temperature and significant
magnetocapacitance above this blocking temperature, as large as 0.4\% at 
$H$ = 10\,kOe.  For both samples the magnetic field induced change in 
dielectric constant was proportional to the square of the sample 
magnetization. These measurements suggest that the dielectric properties of 
magnetic nanoparticles are closely related to the disposition of magnetic 
moments in the system.  As neither bulk \gfo\, nor \mfo\, are magnetoelectric 
materials, this magnetodielectric coupling is believed to arise from extrinsic 
effects which are discussed in light of recent work relating magnetoresistive
and magnetocapacitive behavior. 
\end{abstract}

\pacs{75.50Tt, 77.22.Gm, 77.84.Lf}

\maketitle

Given the renewed activity in investigating materials exhibiting strongly coupled 
electric and magnetic properties,\cite{spaldin} there is great interest
to develop new systems with large magnetocapactive 
couplings. In this contribution, we experimentally explore the
possibility of developing new magnetodielectric materials using magnetic
nanoparticles, and discuss our results in the context of a magnetoresistive
origin for magnetocapacitive couplings.\cite{catalan}

Nanoscale composites can often show large interactions between the magnetic and
electronic properties. For example, in composite materials formed by depositing
CoFe$_2$O$_4$ nanopillars in a BaTiO$_3$ matrix, the elastic coupling between
the different constituents gives rise to sizable magnetoelectric 
coupling.\cite{ramesh} Here we focus on magnetic nanoparticle systems, whose
magnetic properties are well understood in terms of thermally activated spin 
flipping,\cite{brown} and yet have not been much studied from the viewpoint
of the relationship between dielectric and magnetic
properties.\cite{pelster,mallikarjuna,gich}  Very recent work on
$\epsilon$-Fe$_2$O$_3$ nanoparticles show large magnetocapacitive couplings,
which are attributed to intrinsic magnetoelectric couplings in this
material\cite{gich}.  

The systems investigated here are magnetic nanoparticles of \gfo\, and \mfo\, 
embedded in insulating matrices. The magnetic properties of these nanoparticle 
systems are well understood.\cite{mfo,gfo} The \mfo\, nanoparticles were 
prepared through the high-temperature decomposition of metal acetylacetonate
precursors, and are capped with long chain organic molecules, which form the 
insulating matrix.\cite{masala}  The nanoparticles were characterized using 
X-ray diffraction (XRD) and electron microscopy, showing they are crystalline 
with an average diameter of 4.6\,nm.  The \gfo\, nanoparticles were prepared 
through a set of sequential reactions described elsewhere (ref. \cite{buc}). 
XRD characterization show these nanoparticles are crystalline and
electron microscopy indicated an average particle diameter of 5.5\,nm.  

AC magnetic susceptibility measurements were performed on a Quantum Design PPMS 
between $T$ = 2\,K and $T$ = 300\,K at magnetic fields up to $H$ = 10\,Oe. In
order to measure the dielectric properties of these materials,  we cold pressed
approximately 50 mg of sample into a thin, solid pellet. For the \mfo\, sample,
we deposited gold electrodes on opposite sides of this pellet, and attached
platinum leads using silver epoxy. For the \gfo\,  sample electrodes were
fashioned and the leads attached using silver epoxy. We mounted the sample in
the PPMS using a custom designed probe and measured the complex dielectric
constant with measuring frequencies between {\it $\omega_m/2\pi$} = 5 kHz and
{\it $\omega_m/2\pi$} = 1 MHz using an Agilent 4284A LCR meter.

The upper panels of FIG.\,1 show the temperature dependence of the out-of-phase
AC magnetic susceptibility $\chi^{\prime\prime}$ (loss component) 
of the \mfo\, and \gfo\, nanoparticles, and the lower panels show the 
temperature dependence of out-of-phase dielectric loss tangent 
$\epsilon^{\prime\prime}$ recorded at $H$ = 1\,kOe. The data were acquired
at different measurement frequencies.  In the N\'eel-Brown model,\cite{brown}
the characteristic relaxation time for the magnetization of each
(non-interacting) nanoparticle is given by:

\begin{equation} \tau = \tau_0 \exp(E_A/k_B T) \end{equation}

\noindent where $\tau$ is the relaxation time for the magnetic moment, $\tau_0$
is a microscopic timescale (normally close to $10^{-9}$ seconds), $E_A$ is the
energy barrier to moment reversal (taken to be the product of nanoparticle
volume $V$ and magnetocrystalline anisotropy $K$; $E_A = KV $), and $T$ the 
temperature. At high temperatures, $\tau$ will be smaller than the 
characteristic measuring time -- the nanoparticles are in the 
superparamagnetic state -- while at low temperatures $\tau$ will be large, and
the nanoparticle moments are thermally blocked. The peak in 
$\chi^{\prime\prime}$ occurs 
at the temperature where $\tau\omega_m=1$, with $\tau$ determined from 
Eq.\,1, and $\omega_m$  is the AC measuring frequency. 

\begin{figure}[tb] \begin{center} \includegraphics[width=8cm]{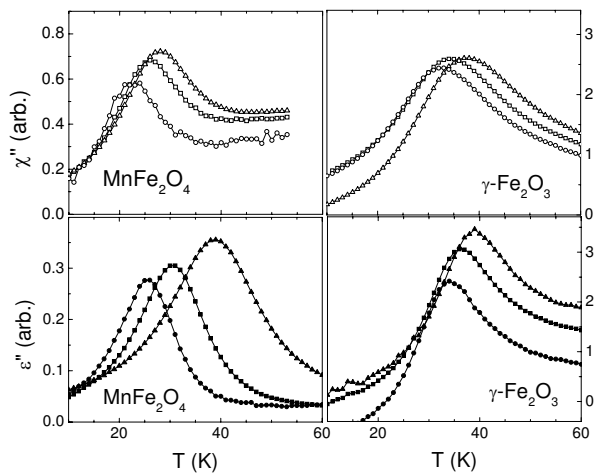}
\caption{Upper left panel: Out-of-phase component of magnetic susceptibility 
of the 4.6\,nm \mfo\, nanoparticles plotted versus temperature at $H$ = 1 kOe. 
The AC measuring frequencies are: 100 Hz (leftmost curve), 
3\,kHz (middle curve), and 10\,kHz (rightmost curve). 
Lower left panel: Out-of-phase dielectric component of the 4.6\,nm \mfo\,  
nanoparticles plotted versus temperature at $H$ = 1 kOe. 
Frequencies: 5\,kHz (leftmost curve), 50\,kHz (middle curve), and 1 MHz 
(rightmost curve). Upper right panel: Out-of-phase component of magnetic 
susceptibility of the 5.5\,nm \gfo\, nanoparticles plotted versus temperature 
at $H$ = 1\,kOe. Frequencies: 1\,kHz (leftmost curve), 3\,kHz (middle curve), 
and 10\,kHz (rightmost curve). 
Lower right panel: Out-of-phase dielectric component 
of the 10\,nm \gfo\, nanoparticles plotted versus temperature at $H$ = 1 kOe. 
Frequencies: 10\,kHz (leftmost curve), 30\,kHz (middle curve), and 100\,kHz  
(rightmost curve).} 
\label{Figure1} \end{center} \end{figure} 

The lower panels in FIG.\,1 show that the loss component of the dielectric
constant of \mfo\, and \gfo\, exhibit peaks at temperatures commensurate with 
the transition from the high temperature superparamagnetic regime to the low
temperature blocked state.  This argues that the dielectric properties of the
\mfo\, and \gfo\, nanocomposites are sensitive to the magnetic dynamics. 
We believe these nanoparticles are rigidly fixed within the insulating matrix 
at all temperatures, so this dielectric anomaly does not arise from any 
physical motion of the nanoparticles. This observation that the peaks in 
$\chi^{\prime\prime}$ and $\epsilon^{\prime\prime}$ in both systems show 
as correspondence is significant because it demonstrates that this coupling 
between magnetic and dielectric properties of nanoparticles is apparently 
general feature of such systems, and does not depend on any specific
material property.

Furthermore, if the underlying mechanism for both the magnetic
and dielectric relaxation is thermally activated moment reversal (Eq.\,1)
in individual
nanoparticles, the magnetic and dielectric data should fall on the same curve. 
Figure 2 shows the Arrhenius curve for the low temperature relaxation feature
for \gfo\, determined using  magnetic and dielectric measurements on the same
plot.  Within experimental error, these two curves agree very well, giving an
activation energy of $E_A$ = 710\,K.  This agreement provides very strong
evidence that the anomalies observed in the dielectric constant of these systems
have their origins in the magnetic properties of the nanoparticles.  

\begin{figure}[tb] \begin{center} \includegraphics[width=6cm]{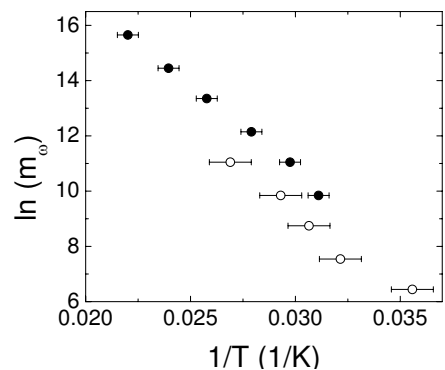}
\caption{Natural logarithm of the measuring frequency plotted against the
inverse temperature of the loss peak for \gfo.  The open symbols were extracted
from the magnetization data, the solid symbols from the dielectric data.}
\label{Figure2} \end{center} \end{figure}

As a final probe of the magnetodielectric properties of these \mfo\, and \gfo\,
systems, we investigated the magnetic field dependence of the dielectric
constant at fixed temperature.  The upper panel of Figure 3 plots the fractional
change in dielectric constant of the \mfo\, composite as a function of magnetic
field.  The dielectric constant of \mfo\, decreases by almost 0.1\% when a
magnetic field of $H$ = 10 kOe is  applied at $T$ = 70\,K.  The lower panel of
Figure 3 shows the relative change in dielectric constant of \gfo\,  as a
function of magnetic field at several different temperatures. Well above  the
blocking temperature, at $T$ = 200\,K and $T$ = 300\,K, the dielectric  constant
increases by over 0.4\% in a field of $H$ = 10 kOe.  However, at  $T$ = 20\,K and
$T$ = 100\,K the dielectric constant is effectively independent  of magnetic
field, at least below $H$ = 10 kOe.  The square of the magnetization is plotted
as the solid symbols in Fig. 4, at $T$ = 70\,K for \mfo, and 
$T$ = 300\,K for \gfo.  For
both these nanoparticle samples, the fractional change of the magnetic field
induced change in the dielectric constant can be well approximated by:

\begin{equation} \frac{\Delta \epsilon}{\epsilon}=\alpha M^2. \end{equation}

\noindent A similar dependence of the change in dielectric constant on the
square of the magnetization has been observed in several bulk
materials.\cite{katsufuji,lawes,kimura}  For \mfo, nanoparticles, $\alpha$ is
small and negative, while for \gfo, $\alpha$ is roughly a factor of 20 larger
and positive.

\begin{figure}[tb] \begin{center} \includegraphics[width=8cm]{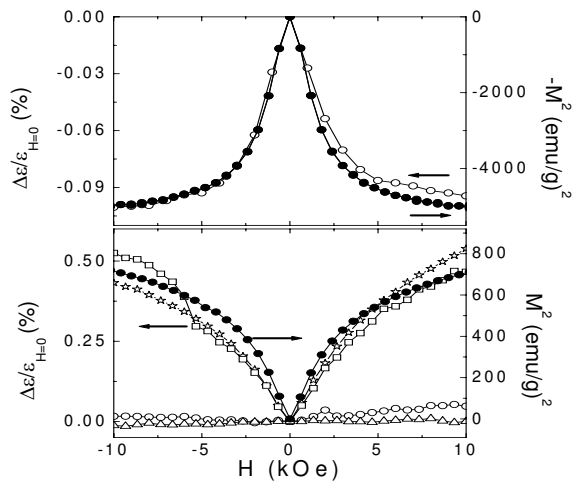}
\caption{Upper Panel:  Relative change in the dielectric constant of the 
4.6\,nm \mfo\, nanoparticle system as a function of external magnetic field, 
measured at $T$ =70\,K and a frequency of 1\,MHz.  Solid symbols plot the 
{\it negative} of the square of the sample magnetization at $T$ = 70\,K.  
Lower Panel: Relative change in the dielectric constant of the 5.5\,nm 
\gfo\, nanoparticle system as a function
of external magnetic field, at temperatures from $T$ = 20\,K to $T$ =300\,K, 
as indicated on the figure. The measuring frequency was 300\,kHz.  The solid 
symbols plot the square of the sample magnetization at $T$ = 300\,K.}
\label{Figure3} \end{center}\end{figure}

These results can be well understood in terms of a recent proposal suggesting
that inhomogeneous systems with large magnetoresistance should also show
significant magnetocapacitive effects.\cite{catalan}  In this framework, the
Maxwell-Wagner capacitor model is used to predict the dielectric response of
magnetoresistive materials.  The measured capacitance depends on the resistance
of the bulk-like and interfacial layers in the system.  If the resistance of
these layers is changed in a magnetic field, the capacitance will change as
well, leading to a magnetodielectric shift. It has been established that
magnetic nanoparticles often show magnetoresistive properties,\cite{tang,liu}
attributed to spin-polarized tunneling, so magnetic nanoparticles
may be expected to show such magnetodielectric effects. Our data agree
qualitatively with several predictions of this model:  the shift in dielectric
constant is proportional to the square of the magnetization, and we observe both
\emph{positive} (\gfo) and \emph{negative} (\mfo) magnetocapacitance.  We can
also compare our results with quantitative predictions of the model.  With the
same assumptions made by Catalan\cite{catalan} together with published values
for the magnetoresistance of \gfo\, (-2\% magnetoresistance at 
$H$ = 10\,kOe\cite{tang} the predicted magnetocapacitance for \gfo\, at 
300\,K is on the order of 1\%. This is comparable to our observed value 
of 0.4\%. 

In conclusion, we have investigated the dielectric properties of \mfo\, and 
\gfo\, magnetic nanoparticle composites.  Both of these materials exhibit a peak
in dielectric loss at the magnetic blocking temperature, and both show a field
dependent dielectric constant above the blocking temperature which varies as the
square of the sample magnetization.   Our data, including the different signs
for the magnetocapacitance in \gfo\, and \mfo, can be well understood using a
recently proposed model for magnetodielectric coupling in magnetoresistive
systems.\cite{catalan} These results are significant because they provide
experimental confirmation for composite magnetodielectric systems with 
magnetoresistive constituents, with electronic rather than elastic magnetocapacitive couplings. Furthermore, our results emphasize that 
magnetodielectric anomalies in themselves are not proof of
multiferroic behavior.\cite{catalan}

The work at Los Alamos National Laboratory was supported by the LDRD program. 
We acknowledge helpful conversations with A. P. Ramirez.  OM and RS acknowledge
support from the NSF Chemical Bonding Center (Chemical Design of Materials)
under Award No. CHE-0434567, and from the Donors of the American Chemical
Society Petroleum Research Fund.

\end{document}